\documentclass[fleqn,10pt]{wlscirep}
\usepackage[utf8]{inputenc}
\usepackage[T1]{fontenc}
\usepackage{bm}
\newcommand\setcurrentname[1]{\def\@currentlabelname{#1}}
\makeatother
\title{Crow instability of vortex lines in dipolar superfluids}

\title{Crow instability of vortex lines in dipolar superfluids}
\author[1,*]{Srivatsa B. Prasad}
\author[1,+]{Nick G. Parker}
\author[1,2,-]{Andrew W. Baggaley}

\affil[1]{Joint Quantum Centre Durham-Newcastle, School of Mathematics, Statistics and Physics, Newcastle University,
	Newcastle upon Tyne, NE1 7RU, United Kingdom}
\affil[2]{Department of Mathematics and Physics, Lancaster University,
	Lancaster, LA1 4YF, United Kingdom}
\affil[*]{srivatsa.badariprasad@newcastle.ac.uk}
\affil[+]{nick.parker@newcastle.ac.uk}
\affil[-]{a.baggaley1@lancaster.ac.uk}

\begin{abstract}
	In classical inviscid fluids, antiparallel vortices perturbed by Kelvin waves exhibit the Crow instability, where the mutual interaction of the Kelvin modes renders them dynamically unstable. This results in the approach and reconnection of the vortices, leading to a cascaded decay into ever-smaller vortex loops. Through mean-field simulations we study the Crow instability of quantum vortex lines in a superfluid whose atoms are subject to the anisotropic, long-ranged dipole-dipole interaction. We observe that the direction of dipole polarization plays a crucial role in determining the dynamically favored Kelvin modes. The subsequent rate of the instability is linked to the mediation of the vortex curvature by the effective dipole-dipole interaction between the vortices themselves. The vortex curvature is strongly suppressed and modes of lower wavenumber are preferred when the dipole polarization is parallel to the vortices, whereas the curvature is maximized for polarizations along the vortices' separation axis. For polarizations along the binormal axis, modes of higher wavenumber are favorable but the instability rate is considerably inhibited. This paves the way to a deeper understanding of vortex reconnections, vortex loop cascades and turbulence in dipolar superfluids.
\end{abstract}
\begin{document}
	
	\flushbottom
	\maketitle
	\thispagestyle{empty}
	
	\section*{Introduction}
	
\begin{figure}
		\centering
		\includegraphics[width=0.5\linewidth]{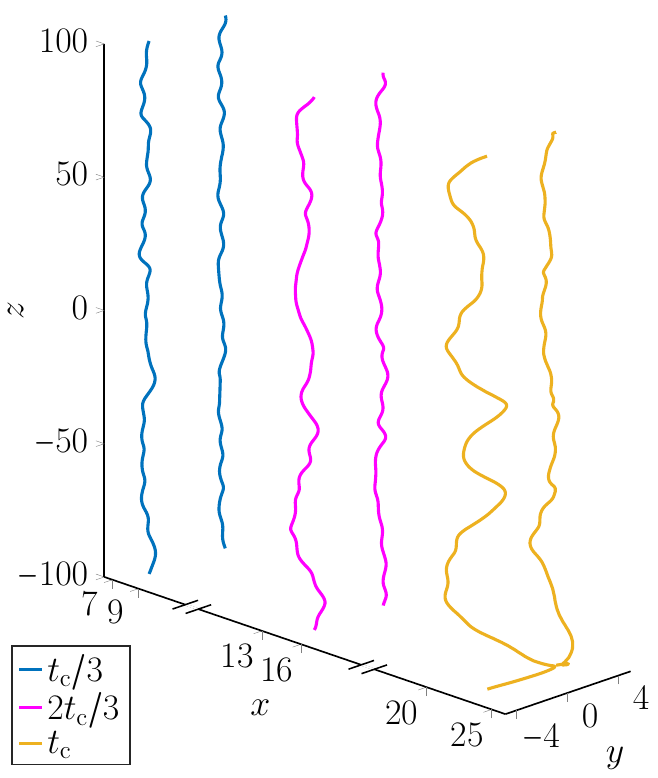}
		\caption{The Crow instability of an antiparallel quantum vortex pair in a nondipolar superfluid. With $t_{\mathrm{c}}$ the time till the first reconnection, the vortices are depicted at the times $t_{\mathrm{c}}/3$, $2t_{\mathrm{c}}/3$ and $t_{\mathrm{c}}$.}
		\label{fig:crowevolutionnondipolar}
	\end{figure}

	Classical inviscid fluids play host to a variety of instabilities arising from the interaction of waves and vortices, and understanding their properties is a central tenet of fluid dynamics. One such example is the celebrated \textit{Crow instability}, where a pair of vortices with antiparallel vorticities is unstable against transverse perturbations that induce helical Kelvin waves along each vortex. These waves grow in amplitude until the vortices reconnect to form a series of vortex loops. The subsequent cascade of reconnections to ever-smaller loops culminates in their dissipation as seen in the suppression of wingtip vortices of aircraft through interactions with contrails~\cite{aiaa_34_47_2172-2179_1970}. The promise of analogues of such instabilities have proven to be a fruitful source of inspiration in the study of quantum fluids, such as Bose-Einstein condensates, that exhibit superfluidity. Unlike classical inviscid fluids, superfluids can flow without viscous dissipation and are characterized by a superfluid order parameter~\cite{pitaevskiistringaribec}. The phase coherence of this order parameter ensures that the vorticity of a superfluid can only be nonzero along discrete, infinitesimally thin lines. Each \textit{quantum vortex} line boasts a quantized circulation and is a node of the superfluid density~\cite{rmp_81_2_647-691_2009}. As such, dynamical processes involving quantum vortices, such as reconnections and annihilations,~\cite{prl_71_9_1375-1378_1993, physfluids_24_1_125108_2012} cannot change the total circulation as it is a topologically conserved quantity~\cite{rmp_81_2_647-691_2009}. Whereas the ground state of an ensemble of quantum vortices -- typically, a triangular vortex lattice -- is well-understood~\cite{prl_87_6_060403_2001, prl_87_12_120405_2001, prl_91_11_110402_2003, prl_93_19_190401_2004}, the interwoven phenomena of instabilities, turbulence and disorder of incompressible (vortices) and compressible (phonon) excitations in superfluids are of considerable interest~\cite{prl_103_4_045301_2009, physrep_622_1-52_2016, nature_539_7627_72-75_2016, annrevcondmat_11_37-56_2020, avsquantsci_5_2_025601_2023}. While certain superfluid instabilities such as the snake instability~\cite{pre_51_5_4479-4484_1995} have no classical counterpart it has been shown that superfluids can become turbulent through quantum analogues of classical hydrodynamic instabilities such as the Kelvin-Helmholtz~\cite{pra_97_5_053608_2018, scipostphys_14_2_025_2023, natphys_20_6_939-944_2024}, Rayleigh-Taylor~\cite{pra_80_6_063611_2009, pra_81_5_053616_2010} and Richtmyer-Meshkov~\cite{pra_82_4_043608_2010} instabilities. The Crow instability, too, emerges in pairs of antiparallel quantum vortex lines~\cite{pre_51_5_4479-4484_1995, jphysa_34_47_10057-10066_2001, pra_84_2_021603r_2011, mplb_27_14_1350097_2013, prf_2_4_044701_2017}. This is depicted in Fig.~\ref{fig:crowevolutionnondipolar}, showing how the transverse excitations of the vortex lines grow until the two vortices have approached each other closely enough for a reconnection into loops to occur
	.
	
	Generally, investigations into hydrodynamic instabilities in superfluids have focussed on systems where the interactions between its constituents can be approximated as isotropic and short-ranged. Magnetic dipolar Bose-Einstein condensates (dBECs), comprised of certain lanthanide atoms with large, permanent magnetic dipole moments, are a counterpoint to this paradigm. When these dipole moments are uniformly polarized by an applied magnetic field, a dBEC exhibits a wealth of exotic phenomena arising from a delicate interplay of the long-ranged and anisotropic dipole-dipole interaction (DDI) and the residual short-ranged, effectively isotropic van der Waals repulsion~\cite{repprogphys_72_12_126401_2009, jphyscondesmatter_29_10_103004_2017, repprogphys_86_02_026401_2023}. The foremost consequence of this is the tendency of a dBEC to exhibit anisotropy, whether it be its density profile under external confinement~\cite{pra_71_3_033618_2005, prl_95_15_150406_2005}, the speed of sound and superfluid critical velocity~\cite{prl_106_6_065301_2011, prl_121_3_030401_2018}, or the profile of the core of an embedded quantum vortex~\cite{pra_79_6_063622_2009, prl_111_17_170402_2013, crphys_24_S3_1-20_2023, pra_109_6_063323_2024}. There is a tendency towards stratification and short-ranged order, both of which are often associated with the presence of intermediate-wavelength roton excitations. This results in novel behavior such as non-triangular vortex lattices with density striping~\cite{prl_95_20_200402_2005, prl_95_20_200403_2005, jphyscondesmatter_29_10_103004_2017, pra_98_2_023610_2018, crphys_24_S3_1-20_2023}, density ripples about vortex cores in quasi-two-dimensional dBECs~\cite{prl_111_17_170402_2013}, stratification during turbulence~\cite{prl_121_17_174501_2018}, pattern formation~\cite{pra_86_3_033606_2012, prr_3_3_033125_2021}, and the existence of supersolid phases~\cite{prx_9_1_011051_2019, prx_9_2_021012_2019, prl_122_13_130405_2019}. Theoretical studies of quantum vortices in dBECs have also predicted a substantial influence of the DDI upon the velocities and trajectories of vortex pairs~\cite{prl_111_17_170402_2013, jphysb_47_16_165301_2014, jlowtempphys_204_1-2_1-11_2021, pra_109_6_063323_2024, pra_109_2_023313_2024} and the Kelvin wave spectra of single vortices~\cite{prl_100_24_240403_2008, pra_98_6_063620_2018}. These results, which are now of ever greater relevance after the experimental realization of vortices in both the superfluid and supersolid phases of a dBEC~\cite{natphys_18_12_1453-1458_2022, nature_635_8038_327-331_2024}, suggest that the onset of a hydrodynamic instability such as the Crow instability is sensitive to the magnitude and direction of the dipole moment polarization.
	
	Thus, in this article we present a systematic analysis of the Crow instability in a uniform dBEC and elucidate how the instability depends on the DDI. Initially, we give an overview of the underlying theory used to model perturbed quantum vortices in a uniform, three-dimensional dBEC. We then proceed to introduce a scenario where a pair of antiparallel vortices is subjected to a random perturbation and provide a qualitative discussion of the results of one such simulation, thereby illustrating the influence of the DDI upon the vortices when the Crow instability is triggered. We then proceed to analyse various properties of the vortex lines during their evolution towards reconnection, from the relative populations of different Kelvin modes and the growth rates of Kelvin mode amplitudes to the global line-averaged curvature of the vortices. Finally, these findings are summarized and their implications, as well as some possible generalizations of our investigation, are discussed.
	
	\section*{Formalism}
	
	In this article, we investigate the dynamics of perturbed pairs of quantum vortices in a dBEC through propagating the dipolar Gross-Pitaevskii equation (dGPE) governing the superfluid order parameter, $\psi(\mathbf{r}, t)$. For a system composed of a single atomic species of mass $m$ and magnetic dipole moment $\mu_{\mathrm{d}}$, polarized uniformly by a magnetic field parallel to the unit vector $\mathbf{B}$, the dGPE is given by~\cite{repprogphys_72_12_126401_2009, jphyscondesmatter_29_10_103004_2017, repprogphys_86_02_026401_2023},
	\begin{equation}
		i\hbar\frac{\partial\psi}{\partial t} = \left\lbrace-\frac{\hbar^2}{2m}\nabla^2 + gn + \mu_0\mu_{\mathrm{d}}^2\int\mathrm{d}^3r'\,V_{\mathrm{dd}}(\mathbf{r} - \mathbf{r}')n(\mathbf{r}') - \mu\right\rbrace\psi. \label{eq:dgpe}
	\end{equation}
	The dGPE incorporates both a short-ranged two-body interaction of \textit{positive} strength $g = 4\pi\hbar^2a_{\mathrm{s}}/m$, where $a_{\mathrm{s}}$ represents the scattering length of the atom-atom scattering potential, and the long-ranged magnetic dipole-dipole interaction (DDI). The DDI is defined as~\cite{repprogphys_72_12_126401_2009}
	\begin{equation}
		V_{\mathrm{dd}}(\mathbf{r}) = \frac{1}{4\pi}\left[\frac{1 - 3\left(\mathbf{B}\cdot\hat{\mathbf{r}}\right)^2}{r^3}\right], \label{eq:ddireal}
	\end{equation}
	and the ratio of the interaction strengths of the DDI and the short-ranged interaction is generally represented by the parameter $\varepsilon_{\mathrm{dd}} = m\mu_0\mu_{\mathrm{d}}^2/(12\pi\hbar^2a_{\mathrm{s}})$. Here, $\mu_0$ is the permeability of free space. Additionally, in Eq.~\eqref{eq:dgpe} the superfluid density and chemical potential are represented as $n = |\psi|^2$ and $\mu$, respectively. 
	
	As encapsulated by Eq.~\eqref{eq:ddireal}, the long-ranged DDI between two atoms in a dBEC is dependent on the angle between their separation, $\mathbf{r}$, and $\mathbf{B}$. When this angle, $\arccos(\mathbf{B}\cdot\hat{\mathbf{r}})$, is less (more) than the critical angle $\theta_{\mathrm{c}} = \arccos(1/\sqrt{3}) \approx 54.7\,\mathrm{ deg}$, the DDI is repulsive (attractive). The resulting tendency of the superfluid atoms to realign themselves to minimize their mutual DDI potential energy is why the condensate density can exhibit anisotropy in instances where a nondipolar BEC would be isotropic. When $\varepsilon_{\mathrm{dd}} > 1$, the solutions of Eq.~\eqref{eq:dgpe} are unstable to collapse as the dipolar attraction overwhelms the short-ranged repulsion. From a theoretical viewpoint, this necessitates the inclusion of a beyond-mean-field energy correction~\cite{ijmpb_20_24_1791-1794_2006, pra_86_6_063609_2012, pra_94_3_033619_2016} in such regimes to account for the existence and stability of exotic states such as \textit{quantum droplets}~\cite{nature_539_7628_259-262_2016, prx_6_4_041039_2016} and supersolids~\cite{prx_9_1_011051_2019, prx_9_2_021012_2019, prl_122_13_130405_2019}. Thus, for the sake of simplicity, we focus on the regime $0 \leq \varepsilon_{\mathrm{dd}} < 1$, where no such correction is necessary~\cite{repprogphys_86_02_026401_2023}. This renders it convenient to work in natural units where energies are scaled by $\mu_{\mathrm{g}} = 4\pi\hbar^2a_{\mathrm{s}}(1 - \varepsilon_{\mathrm{dd}})n_0/m$, the chemical potential of a dBEC of uniform background density $n_0$. Simultaneously, times and lengths are scaled by $\tau = \hbar/\mu_{\mathrm{g}}$ and the healing length $\xi=\hbar/\sqrt{m\mu_{\mathrm{g}}}$, respectively, while the order parameter is scaled by $\sqrt{n_0}$~\cite{primer}. Rendering Eq.~\eqref{eq:dgpe} in this manner results in a description of the dBEC independent of parameters such as $m$, $a_{\mathrm{s}}$, or the total atom number, with $\mathbf{B}$ and $\varepsilon_{\mathrm{dd}}$ being the only remaining free parameters:
	\begin{equation}
		i\frac{\partial\psi}{\partial t} = \left\lbrace-\frac{\nabla^2}{2} + \frac{1}{1 - \varepsilon_{\mathrm{dd}}}\left[n + 3\varepsilon_{\mathrm{dd}}\int\mathrm{d}^3r'\,V_{\mathrm{dd}}(\mathbf{r} - \mathbf{r}')n(\mathbf{r}')\right] - 1\right\rbrace\psi. \label{eq:dgpescaled}
	\end{equation}
	
	\section*{Results}
	
	\subsection*{Vortex line profiles: qualitative features}
	
	In a realistic scenario, the Crow instability may be triggered in a trapped Bose-Einstein condensate when the trapping potential landscape is locally inhomogeneous in the vicinity of a pair of antiparallel vortices, thereby resulting in background density gradients that perturb the vortices and seed Kelvin waves.~\cite{pra_84_2_021603r_2011, mplb_27_14_1350097_2013}. In this article we consider an idealized scenario where both vortex lines are already subject to a random transverse perturbation in the initial conditions of the numerical simulations~\cite{pre_51_5_4479-4484_1995, jphysa_34_47_10057-10066_2001, prf_2_4_044701_2017}. Since there is no inhomogeneous or anisotropic trapping applied to the superfluid, we can assume without loss of generality that the unperturbed vortices are (anti-)parallel to the $z$-axis and separated along the $y$-axis. As described in the Methods, \textit{cf.} Eqs.~\eqref{eq:weissmcwilliamsphase}-\eqref{eq:kelvininitconds}, each vortex is initially imprinted with the $40$ lowest-lying Kelvin modes of the computational domain with a random phase, $\eta_q$, associated with each mode $q$. We have populated an ensemble of 6 distinct sets of $\lbrace\eta_q\rbrace$ with each value sampled from a uniform distribution over $[0, 2\pi)$. For each choice of parameters, our analysis of the evolution of the vortex pair involves a set of simulations over this ensemble. The two-body interaction strength ratio $\varepsilon_{\mathrm{dd}}$ assumes one of the values $\lbrace 0, 0.1, \mathellipsis 0.8, 0.9\rbrace$, ranging from the nondipolar limit to just below the maximal threshold, $\varepsilon_{\mathrm{dd}} = 1$, for the validity of mean-field theory. Given the initial vortex configuration, we consider \textbf{B}, the dipole polarization axis, to be parallel to either the vortex lines' translational velocity ($\hat{x}$), their mutual separation ($\hat{y}$), or the lines themselves ($\hat{z}$). In addition, until the final stages of the reconnection process, we desire that the vortex dynamics are driven principally by Kelvin wave–sound interactions rather than the mutual overlap of the vortices. Thus, all of the simulations feature an initial vortex separation of $d = 6.25\xi$. While this choice is somewhat arbitrary, it ensures that the vortex cores do not overlap at $t = 0$ and is also small enough that the reconnection process occurs over a duration that is computationally feasible to simulate.
	
\begin{figure}
		\centering
		\includegraphics[width=\textwidth]{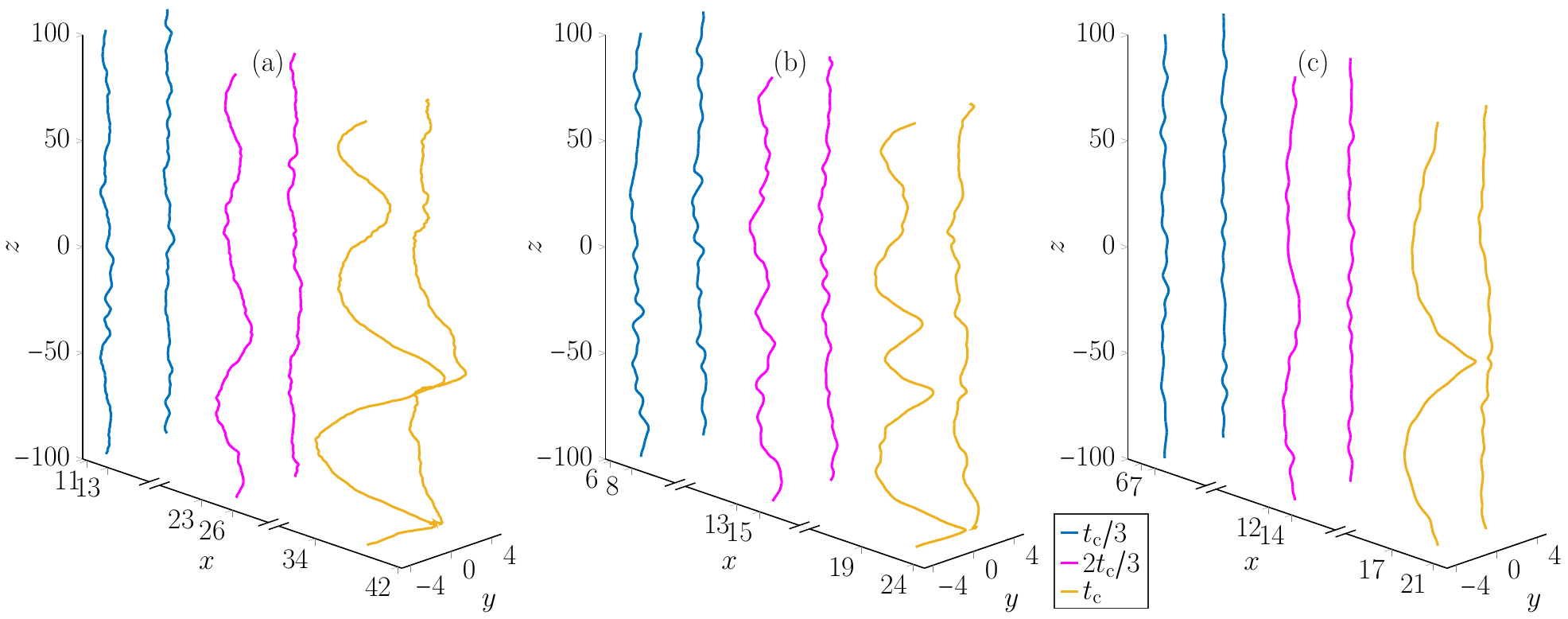}
		\caption{The Crow instability of an antiparallel quantum vortex pair, initially separated along the $y$-axis, in a dipolar superfluid with $\varepsilon_{\mathrm{dd}} = 0.9$ and $\mathbf{B}$ parallel to $\hat{x}$ (a), $\hat{y}$ (b) and $\hat{z}$ (c). With $t_{\mathrm{c}}$ the time till the first reconnection, the vortices are depicted at the times $t_{\mathrm{c}}/3$, $2t_{\mathrm{c}}/3$ and $t_{\mathrm{c}}$.}
		\label{fig:crowevolutiondipolar}
	\end{figure}

	We now turn to describing the qualitative features of the vortex lines as they approach a reconnection. In units of $\tau = \hbar/\mu_{\mathrm{g}}$, the vortex line profiles are computed from the numerical values of $\psi$ at time intervals of $\Delta t = 0.5$ and a reconnection is identified topologically through the change in vortex line terminals from one pair to the next. For a given simulation, we define the last observed time before the reconnection occurs as $t_{\mathrm{c}}$ and therefore the exact reconnection time lies in the interval $(t_{\mathrm{c}}, t_{\mathrm{c}} + \Delta t)$. Our analysis of the vortex lines does not extend past $t = t_{\mathrm{c}}$, such that we restrict our focus to the duration before the disconnected vortex lines have become loops. Returning to Fig.~\ref{fig:crowevolutionnondipolar}, which depicted the Crow instability in a nondipolar superfluid, we note that the snapshots of the vortex lines were taken at the times $t = t_{\mathrm{c}}/3$, $t = 2t_{\mathrm{c}}/3$, and $t = t_{\mathrm{c}}$. Figure~\ref{fig:crowevolutiondipolar} depicts the corresponding snapshots of the vortex pair subjected to the same Kelvin wave initial conditions but with $\varepsilon_{\mathrm{dd}} = 0.9$ and $\mathbf{B}$ being parallel to $\hat{x}$ (a), $\hat{y}$ (b), or $\hat{z}$ (c). Let us stress that the time to the first reconnection, $t_{\mathrm{c}}$, is dependent on $\mathbf{B}$ and $\varepsilon_{\mathrm{dd}}$ as well as the initial Kelvin wave profile. Similarly, the displacement of the vortices from their initial position is dependent on $\mathbf{B}$ and $\varepsilon_{\mathrm{dd}}$ as expected from studies of dipolar superfluid vortex dynamics in the absence of Kelvin waves~\cite{pra_109_6_063323_2024}.
	
	Figure \ref{fig:crowevolutiondipolar} visibly demonstrates a dependence of the Crow instability of the vortex lines on the DDI. Compared to the nondipolar case (Fig.~\ref{fig:crowevolutionnondipolar}), the vortex lines in Fig.~\ref{fig:crowevolutiondipolar} (c) ($\mathbf{B} \parallel \hat{z}$) are noticeably straighter. Whereas the nondipolar limit is characterized by a highly agitated pair of vortex lines when $t = t_{\mathrm{c}}$, narrowly separated at multiple locations along the $z$-axis, only one such point is evident in the $z$-polarized case. This is sharply contrasted by what occurs when $\mathbf{B} \parallel \hat{y}$, the separation axis of the vortices (Fig.~\ref{fig:crowevolutiondipolar} (b)). Here, the three vortex separation local minima are roughly at the same points along the $z$-axis as those of the nondipolar limit and, of the three, the global minimum where the first reconnection occurs is closest to that of the nondipolar case. However, the vortices are closer to each other at the other two minima than when $\varepsilon_{\mathrm{dd}} = 0$; the vortex separations at the upper and lower minima in Fig.~\ref{fig:crowevolutiondipolar} (b) are $3.15\xi$ and $2.82\xi$, respectively, compared to $3.45\xi$ and $3.40\xi$, respectively for the corresponding minima in Fig.~\ref{fig:crowevolutionnondipolar}. This indicates that the second and third reconnections occur earlier for nonzero $\varepsilon_{\mathrm{dd}}$ when $\mathbf{B} \parallel \hat{y}$ than in the nondipolar limit. By contrast, the chief characteristic of the vortex lines in Fig.~\ref{fig:crowevolutiondipolar} (a) ($\mathbf{B}\parallel\hat{x}$) is that the vortices are perturbed more strongly along the $x$-axis at each local minimum of their separation than in the other regimes; this is particularly evident when $t = t_{\mathrm{c}}$. We also note that we have verified these qualitative aspects of vortex pair evolution for smaller values of $\varepsilon_{\mathrm{dd}}$ and thus believe that they are independent of $\varepsilon_{\mathrm{dd}}$.
	
	\subsection*{Vortex line profiles: analysis}
	
\begin{figure}
		\centering
		\includegraphics[width=\textwidth]{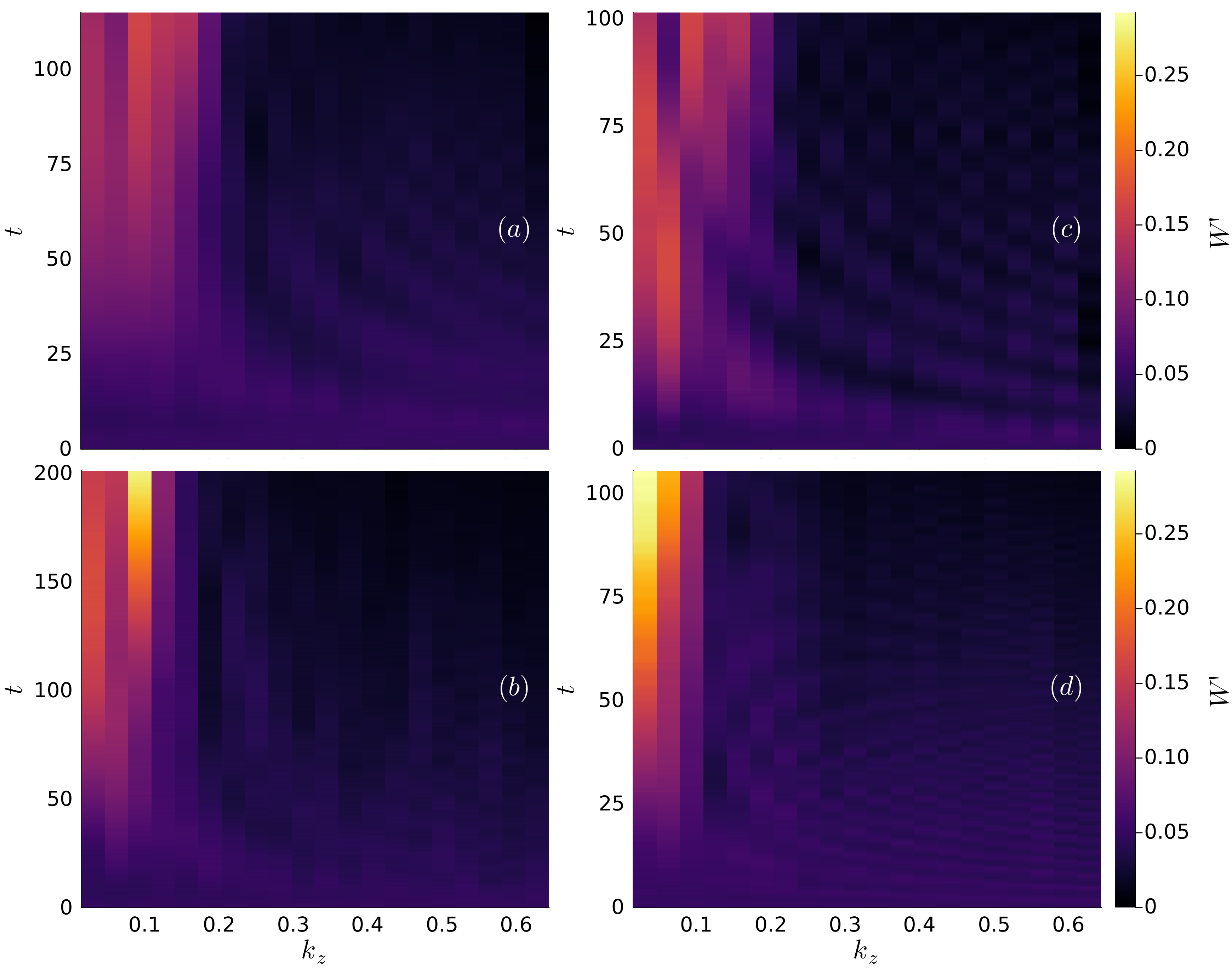}
		\caption{The relative occupation of the lowest-lying Kelvin modes with positive $k_z$ during time evolution, ensemble-averaged over the initial conditions. In (a), $\varepsilon_{\mathrm{dd}} = 0$ and in the remaining subfigures, $\varepsilon_{\mathrm{dd}} = 0.9$; $\mathbf{B}$ is parallel to $\hat{x}$ in (b), $\hat{y}$ (c) and $\hat{z}$ (d).}
		\label{fig:modeaveragedheatmaps}
	\end{figure}

	The significant variation in the vortex line profiles at $t = t_{\mathrm{c}}$, modulated by the DDI, suggests that the relative contributions of the various Kelvin modes on the vortices are modulated by the dipolar parameters $\mathbf{B}$ and $\varepsilon_{\mathrm{dd}}$. This warrants an investigation of the spectral properties of the vortex line profiles, through which one can characterize the time-dependent populations of the Kelvin waves. Indexing the vortices as $n \in \lbrace 1, 2\rbrace$, we write the time-dependent amplitude of the mode on vortex $n$ with wavenumber $k_z$ as $W_n(k_z, t)$; a method for computing these amplitudes from the vortex line profiles is described in the Methods. Initially, we study the time-dependence of the respective contributions of each mode to the overall mode population. The parity symmetry of the initial conditions, $|W_1(k_z, t = 0)| = |W_2(-k_z, t = 0)|$, motivates the definition of the relative population of a Kelvin wave of positive $k_z$ as
	\begin{equation}
		W'(k_z, t) = \frac{|W_1(k_z, t)| + |W_2(-k_z, t)|}{\sum_{k_z > 0}\left[|W_1(k_z, t)| + |W_2(-k_z, t)|\right]}. \label{eq:propamp}
	\end{equation}
	Figure~\ref{fig:modeaveragedheatmaps} depicts $W'(k_z, t)$ in the nondipolar limit (a) as well as the maximally dipolar regime, $\varepsilon_{\mathrm{dd}} = 0.9$, with $\mathbf{B}$ parallel to $\hat{x}$ (b), $\hat{y}$ (c), or $\hat{z}$ (d). In each subplot, we average the results over the ensemble of initial conditions until the minimal value of $t_{\mathrm{c}}$ in the ensemble. This accounts for the distinct endpoints of $t$ in each subplot.
	
	Irrespective of the dipolar parameters, Fig~\ref{fig:modeaveragedheatmaps} demonstrates that the dynamics of the vortices at early times are characterized by the mutual interaction of the modes driving fluctuations of their relative populations. This is accompanied by the suppression of strongly energetic high order modes with $|q| \gtrsim 7$. We note that this justifies the initial mode cutoff, $|q| \leq 20$, as being sufficiently large as to not exert an artificial influence upon the dynamics. It is not until $t \gtrsim 2t_{\mathrm{c}}/3$ and the vortices are nearing their first reconnection that unambiguous signatures of a dependence on the dipolar parameters becomes evident, \textit{cf.} Figs.~\ref{fig:crowevolutionnondipolar} and \ref{fig:crowevolutiondipolar}. In Fig.~\ref{fig:modeaveragedheatmaps} (d), where $\varepsilon_{\mathrm{dd}} = 0.9$ and the dipole polarization is (anti-)parallel to the vortex lines, the Kelvin mode populations monotonically decrease as a function of $k_z$ as $t \rightarrow t_{\mathrm{c}}$. This is consistent with the features of the vortex lines in Fig.~\ref{fig:crowevolutiondipolar} (c) since, at $t = t_{\mathrm{c}}$, each vortex exhibits only one antinode of its displacement from the unperturbed mean. By contrast, polarizing the dipole moments orthogonal to the vorticity of either vortex stimulates the population of higher Kelvin modes. Figure~\ref{fig:modeaveragedheatmaps} (b) illustrates clearly that when $\mathbf{B}$ is parallel to the vortices' velocity, i.e. $\hat{x}$, the $q = 3$ mode is by far the most strongly excited at late times before the first reconnection. Again, this is consistent with the qualitative features at $t = t_{\mathrm{c}}$ in Fig.~\ref{fig:crowevolutiondipolar} (a) where $3$ distinct antinodes are distinguishable on each vortex. Intriguingly, Fig.~\ref{fig:modeaveragedheatmaps} (c) depicts that while modes of odd $q$ are preferentially occupied as $t \rightarrow t_{\mathrm{c}}$ when the dipole polarization is parallel to the separation axis, $\hat{y}$, no single mode is overwhelmingly dominant. Nonetheless, the first $6$ modes contribute to the majority of the mode population just like the nondipolar limit. Indeed, while the two vortex antinodes at $z \approx -20,\,-50$ when $t = t_{\mathrm{c}}$ appear to be closer to instigating reconnections in Fig.~\ref{fig:crowevolutiondipolar} (b) than the corresponding ones are in Fig.~\ref{fig:crowevolutionnondipolar}, the nondipolar vortices are otherwise much more similar to the dipolar vortices in Fig.~\ref{fig:crowevolutiondipolar} (b) than those in (a) and (c).
	
\begin{figure}
		\centering
		\includegraphics[width=0.5\linewidth]{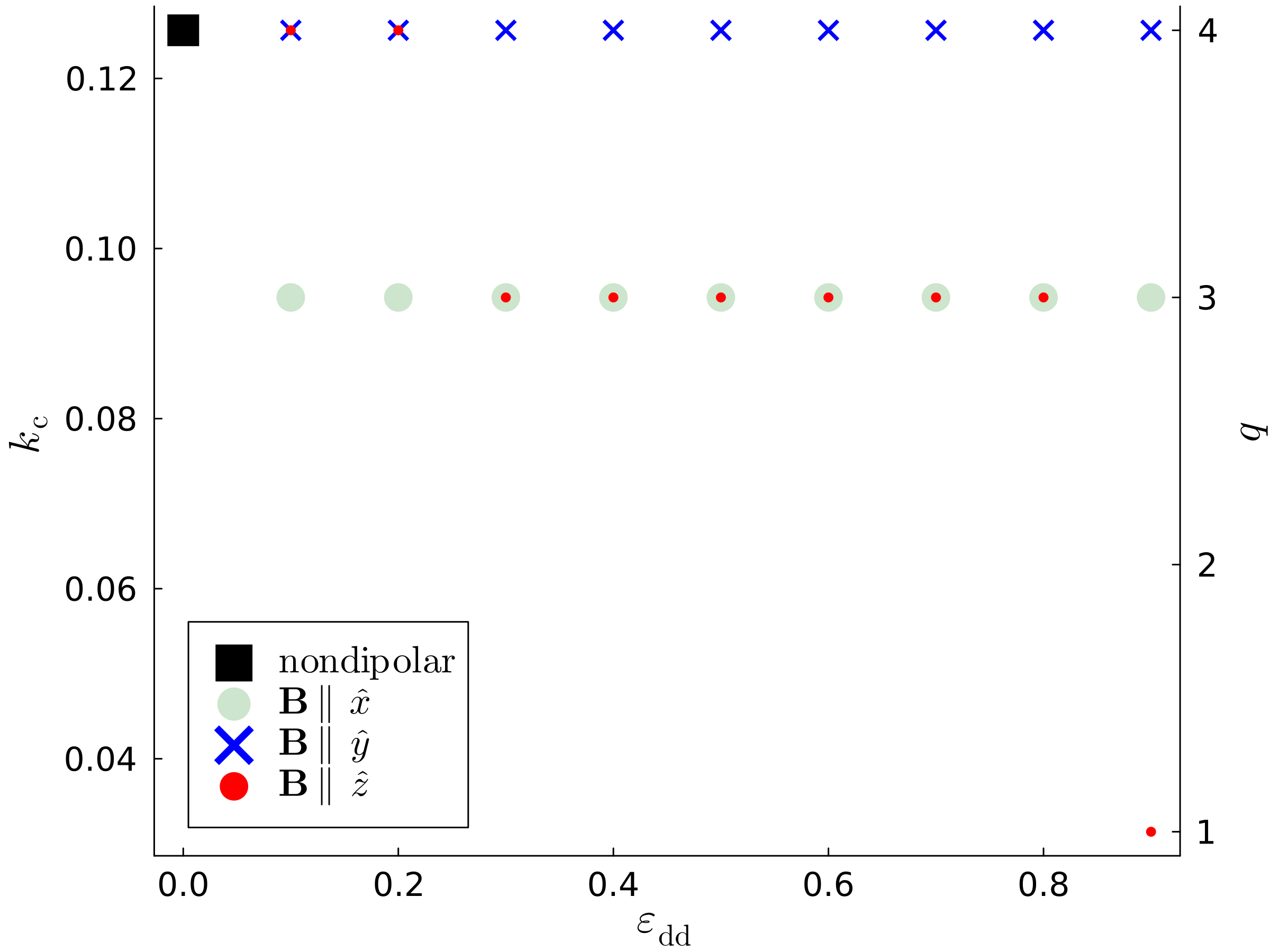}
		\caption{The wavenumber, $k_{\mathrm{c}}$, of the Kelvin wave maximally occupied at $t = t_{\mathrm{c}}$, represented as a mode over the ensemble of initial conditions as a function of $\varepsilon_{\mathrm{dd}}$. The distinct markers correspond to polarizations parallel to $\hat{x}$, $\hat{y}$, and $\hat{z}$, each of which converge in the limit $\varepsilon_{\mathrm{dd}} = 0$ to the nondipolar value of $k_{\mathrm{c}}$; this is represented by a distinct marker. The index of the Kelvin modes, $q$, is also represented for reference.}
		\label{fig:maxmodeeachB}
	\end{figure}

	With these insights on the evolution of the modes in the maximally dipolar regime, $\varepsilon_{\mathrm{dd}} = 0.9$, we proceed to study specific features of the mode population over the full range $\varepsilon_{\mathrm{dd}} \in [0, 0.9]$. One such measure is $k_z \equiv k_{\mathrm{c}}$, the wavenumber that maximizes $W'(k_z, t_{\mathrm{c}})$ and is thus the most strongly excited Kelvin wave at $t = t_{\mathrm{c}}$. Figure~\ref{fig:maxmodeeachB} plots the ensemble mode of $k_{\mathrm{c}}$ as a function of $\varepsilon_{\mathrm{dd}}$ with the values for each $\mathbf{B}$ being represented by distinct markers. For the sake of clarity, we have augmented the $k_{\mathrm{c}}$-axis with the corresponding values of the Kelvin wave index $q$. It is immediately apparent that in the nondipolar limit the $q = 4$ mode is most frequently the dominant one. This corresponds to a wavenumber $k_{\mathrm{c}}\xi = 8\pi/200 \approx 0.126$, which is broadly consistent with predictions in earlier studies of the nondipolar Crow instability that $dk_{\mathrm{c}} \sim 1$~\cite{jphysa_34_47_10057-10066_2001}. When $\varepsilon_{\mathrm{dd}} \neq 0$ the behavior of $k_{\mathrm{c}}$ is directionally dependent on the dipole polarization. When $\mathbf{B} \parallel \hat{y}$, $k_{\mathrm{c}}$ is insensitive to $\varepsilon_{\mathrm{dd}}$ and its ensemble mode is equivalent to the nondipolar value across the full range of $\varepsilon_{\mathrm{dd}}$. This is consistent with the qualitative similarity of the plots of the relative Kelvin mode populations for $\varepsilon_{\mathrm{dd}} = 0$ and $\lbrace\varepsilon_{\mathrm{dd}} = 0.9,\,\mathbf{B}\parallel\hat{y}\rbrace$ in Figs.~\ref{fig:modeaveragedheatmaps} (a) and (c), respectively. However, such robustness with respect to $\varepsilon_{\mathrm{dd}}$ is not evident for dipole polarizations along either $\hat{x}$ or $\hat{z}$. When $\varepsilon_{\mathrm{dd}} = 0.9$ and $\mathbf{B} \parallel \hat{z}$, Fig.~\ref{fig:modeaveragedheatmaps} (d) shows that the $q = 1$ mode is preferentially occupied most often as $t \rightarrow t_{\mathrm{c}}$. This substantial discrepancy away from the nondipolar limit for large $\varepsilon_{\mathrm{dd}}$ is supported by Fig. \ref{fig:maxmodeeachB} where $k_{\mathrm{c}}$ also broadly decreases with increasing $\varepsilon_{\mathrm{dd}}$. Thus, while we do not focus on the dynamics of the vortices for $t > t_{\mathrm{c}}$, the vortex loops formed as a consequence of the Crow instability are unambiguously larger for increasing $\varepsilon_{\mathrm{dd}}$ when the dipole polarization is parallel to the initial vortex configuration. A similar scenario occurs when $\mathbf{B} \parallel \hat{x}$ with the $q = 3$ mode being dominant most frequently for $\varepsilon_{\mathrm{dd}} \geq 0.1$.
	
\begin{figure}
		\centering
		\includegraphics[width=0.4975\linewidth]{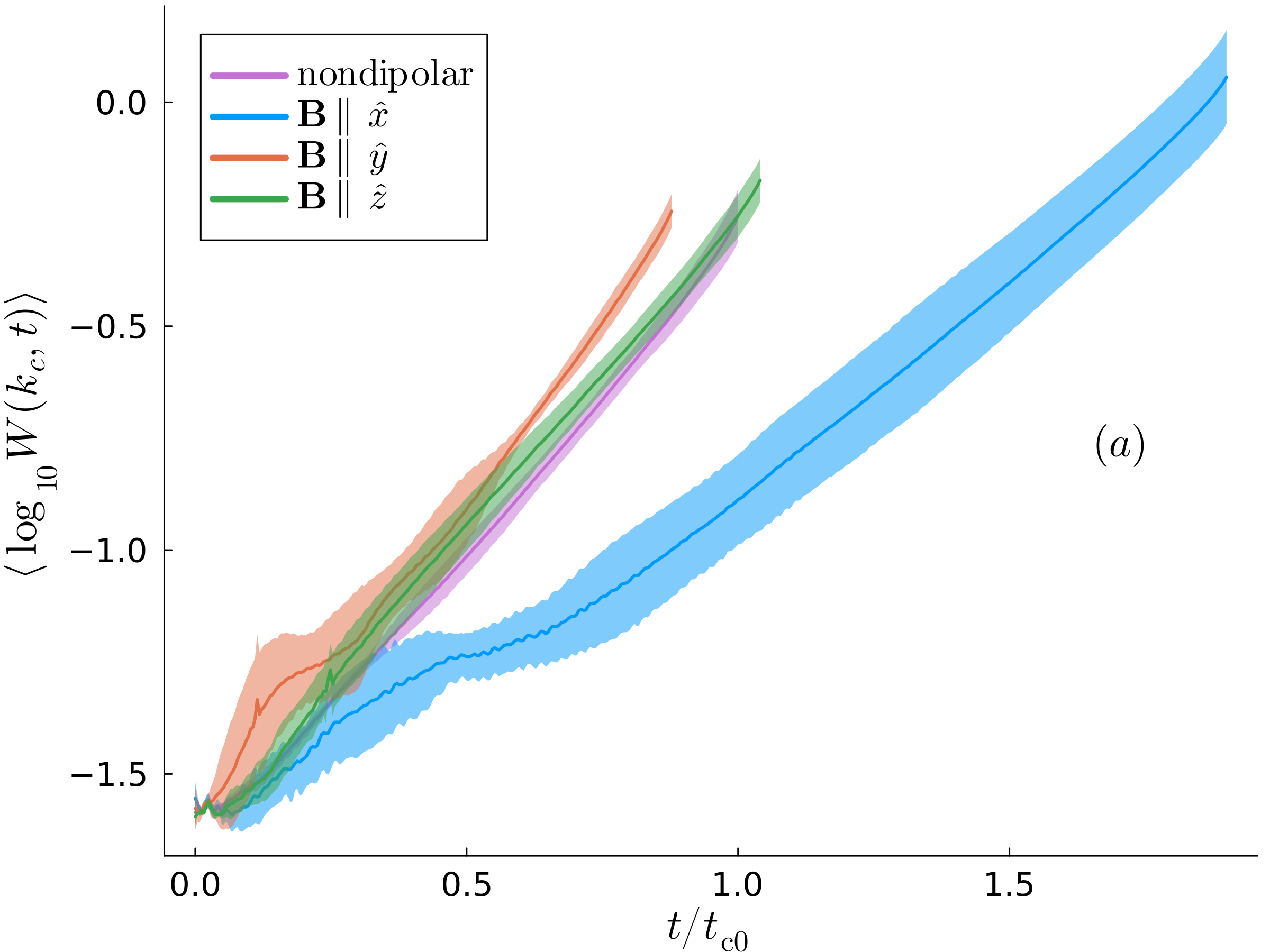}
		\includegraphics[width=0.4975\linewidth]{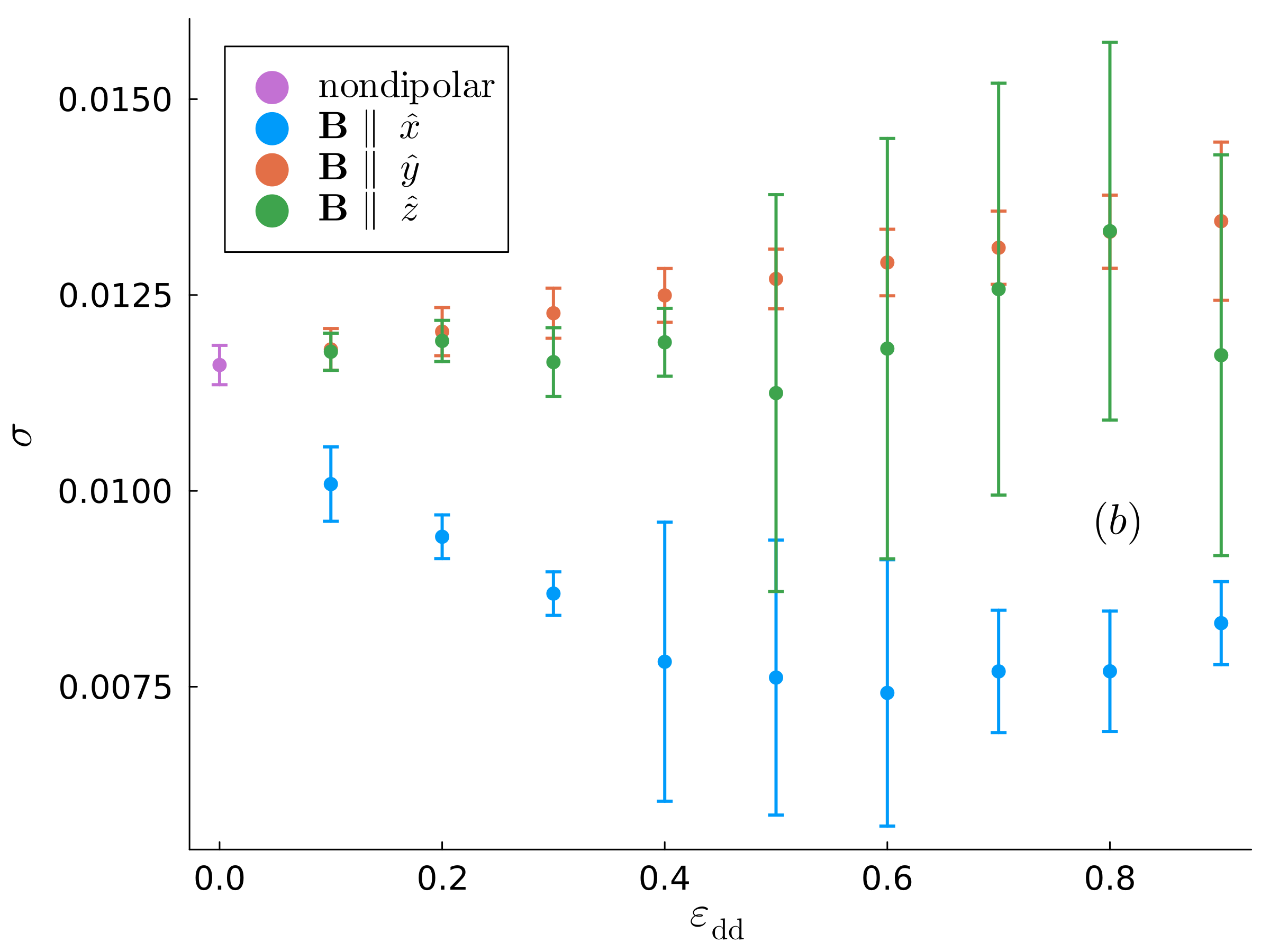}
		\caption{(a) $\log_{10}\,W(k_{\mathrm{c}}, t)$ as a function of $t$, with $k_{\mathrm{c}}$ the maximally unstable mode wavenumber and $\varepsilon_{\mathrm{dd}} = 0.9$. The initial conditions correspond to those in Figs.~\ref{fig:crowevolutionnondipolar} and \ref{fig:crowevolutiondipolar}.  (b) The ensemble-average of $\sigma$, the exponential growth rate of $W(k_{\mathrm{c}}, t)$, as a function of $\varepsilon_{\mathrm{dd}}$. In both plots, $\mathbf{B}\parallel\lbrace\hat{x}, \hat{y}, \hat{z}\rbrace$, and their convergence to our nondipolar results is represented by a distinct marker.}
		\label{fig:modeexponentensembledandmodegrowth}
	\end{figure}

	Given the maximally unstable Kelvin mode, $k_{\mathrm{c}}$, we can also study the evolution of its amplitude, $W(k_{\mathrm{c}}, t)$, over the interval $t \in [0, t_{\mathrm{c}}]$. The early stages of vortex pair evolution are marked by a transient in $W(k_{\mathrm{c}}, t)$ arising from nonlinear interactions with the other modes. This can be attributed to the initially imprinted superfluid phase, defined in Eqs.~\eqref{eq:weissmcwilliamsphase} and \eqref{eq:kelvininitconds} in the Methods, being only an approximation of the true phase of a superposition of Kelvin modes on a pair of antiparallel vortex lines. Thus, a redistribution of Kelvin mode amplitudes occurs as the vortices interact with the background superfluid to establish a valid phase profile. Subsequently, $W(k_{\mathrm{c}}, t)$ grows approximately exponentially in time, i.e. $W(k_{\mathrm{c}}, t) \sim \exp(\sigma t)$ with $\sigma$ the growth rate. Let us define the timescale $t_{\mathrm{c}0}$ as the ensemble-averaged value of $t_{\mathrm{c}}$ in the nondipolar limit. Figure \ref{fig:modeexponentensembledandmodegrowth} (a) plots ensemble-averages of $\log_{10}\,W(k_{\mathrm{c}}, t)$ as a function of $t/t_{\mathrm{c}0}$ from $t = 0$ till the ensemble-averaged value of $t_{\mathrm{c}}/t_{\mathrm{c}0}$; this is presented for $\varepsilon_{\mathrm{dd}} = 0.9,\,\mathbf{B}\parallel\lbrace\hat{x},\hat{y},\hat{z}\rbrace$ as well as the nondipolar limit. The exponential growth of $W(k_{\mathrm{c}}, t)$ with respect to $t$ means that each curve is approximately linear; this is consistent with the Crow instability manifesting itself as a linear dynamical instability of the $k = k_{\mathrm{c}}$ mode. Figure~\ref{fig:modeexponentensembledandmodegrowth} (a) also emphasizes the pronounced dependence of $t_{\mathrm{c}}$ on the dipole polarization, with prior investigations additionally demonstrating the dipolar dependence of the vortex pair's velocity~\cite{pra_109_6_063323_2024}. Extracting the growth rate, $\sigma$, requires excising of the effects of the initial transient in $\psi$ before applying linear regression to $\log_{10}\,W(k_{\mathrm{c}}, t)$. Thus we conduct regression only in the linear growth regime which we take to be the interval $t \in [t_{\mathrm{c}}/2, t_{\mathrm{c}}]$. This yields Fig.~\ref{fig:modeexponentensembledandmodegrowth} (b), where the ensemble average of $\sigma$ is plotted as a function of $\varepsilon_{\mathrm{dd}}$ with distinct markers for $\mathbf{B}\parallel\lbrace\hat{x},\hat{y},\hat{z}\rbrace$. First, we compare our nondipolar result, $\sigma \approx 0.0118$, to pre-existing predictions obtained through linear stability analysis of antiparallel vortices in superfluids~\cite{pre_51_5_4479-4484_1995, jphysa_34_47_10057-10066_2001},
	\begin{equation}
		\sigma(k_{\mathrm{c}})^2 \sim d^{-2}k_{\mathrm{c}}^2\left[\ln(\sqrt{2}d) + 0.38\right],
	\end{equation}
	yielding a prediction $\sigma \approx 0.04$. While both values are of the same order of magnitude, we believe that the discrepancy is mainly due to the nonlinear growth of $W(k_{\mathrm{c}}, t)$ evident in Fig.~\ref{fig:modeexponentensembledandmodegrowth} (a) and the aforementioned reasons for this nonlinearity. When $\varepsilon_{\mathrm{dd}}$ is nonzero, Fig.~\ref{fig:modeexponentensembledandmodegrowth} (b) shows that $\sigma$ is always smaller for $\mathbf{B}\parallel\hat{x}$ than for the other two polarizations and that the $\varepsilon_{\mathrm{dd}} \rightarrow 1$ limit is characterized by the hierarchy $\sigma(\hat{x}) < \sigma(\hat{z}) < \sigma(\hat{y})$. Given that we always specify an initial vortex separation $d = 6.25\xi$, one would expect longer reconnection times for smaller $\sigma$. As such, the observed hierarchy of $\sigma$ in Fig.~\ref{fig:modeexponentensembledandmodegrowth} (b) is consistent with the ensemble-averaged reconnection times, $t_{\mathrm{c}}(\hat{x}) > t_{\mathrm{c}}(\hat{z}) > t_{\mathrm{c}}(\hat{y})$, seen in Fig.~\ref{fig:crowevolutiondipolar} in the regime $\varepsilon_{\mathrm{dd}} = 0.9$.
	
\begin{figure}
		\centering
		\includegraphics[width=0.5\linewidth]{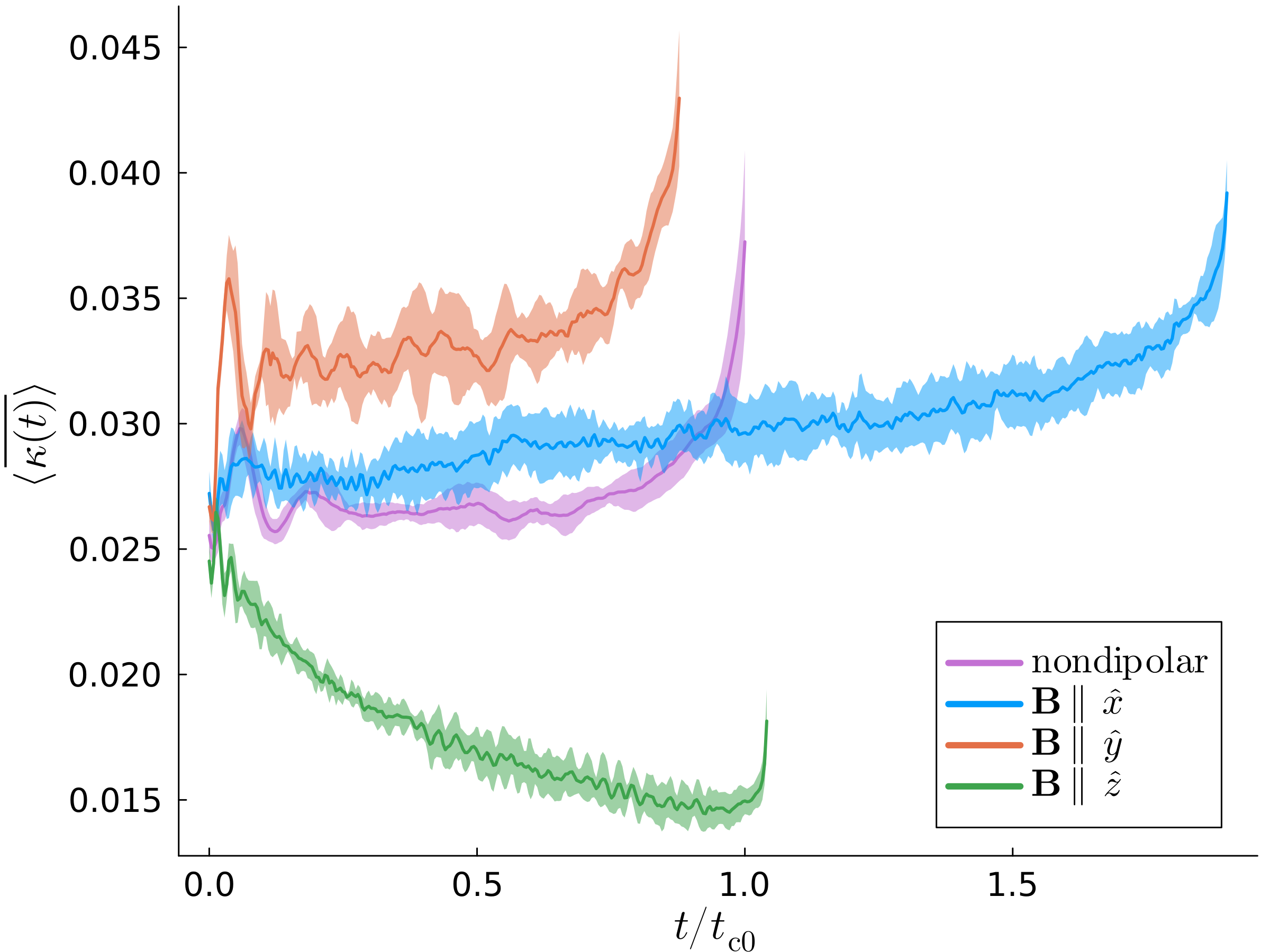}
		\caption{The mean curvature of the vortices ensemble-averaged over the initial conditions, $\langle\overline{\kappa(t)}\rangle$, as a function of $t/t_{\mathrm{c}0}$. Except for the nondipolar case,  we have $\varepsilon_{\mathrm{dd}} = 0.9$.}
		\label{fig:curvensavgtdep}
	\end{figure}

	While the results presented so far focused on the spectral properties of the vortex lines, the dipolar dependence of the maximally occupied Kelvin mode wavenumber, $k_{\mathrm{c}}$, indicates that local geometric properties of the vortex lines are also sensitive to the DDI. One such quantity is the vortices' \textit{curvature}. Let us write the coordinates of the $n$th vortex line at time $t$ as $\bm{\gamma}^{(n)}(s, t) = [\gamma_x^{(n)}(s, t), \gamma_y^{(n)}(s, t), s]$. Suppressing the index $n$ for notational simplicity, the curvature of this line is then defined as~\cite{jfm_808_641-667_2016}
	\begin{equation}
		\kappa(s, t) = \frac{|\partial_s\bm{\gamma}(s, t)\times\partial_{ss}^2\bm{\gamma}(s, t)|}{|\partial_s\bm{\gamma}(s, t)|^3}. \label{eq:curvaturedefn}
	\end{equation}
	Initially, we study the time-dependence of the vortices' curvature, focussing on the maximally dipolar case $\varepsilon_{\mathrm{dd}} = 0.9$. Comparisons can then be made between the vortices' curvature for different dipole polarizations and with the nondipolar limit, as well as with the corresponding results for the evolution of $W(k_{\mathrm{c}}, t)$ in Fig.~\ref{fig:modeexponentensembledandmodegrowth} (a). Let us define $\overline{\kappa(t)}$ as the global curvature at time $t$ found by averaging $\kappa(s, t)$ over $s$ for each vortex and finding the mean of the two spatial averages. We also define $\langle\overline{\kappa(t)}\rangle$ as the ensemble average of $\overline{\kappa(t)}$. Figure \ref{fig:curvensavgtdep} plots $\langle\overline{\kappa(t)}\rangle$ as a function of $t/t_{\mathrm{c}0}$; here, the reader is reminded that $t_{\mathrm{c}0}$ is the ensemble-averaged nondipolar value of $t_{\mathrm{c}}$. A fascinating contrast emerges when examining the features of Fig.~\ref{fig:curvensavgtdep}. Firstly, a universal property of $\langle\overline{\kappa(t)}\rangle$ is that it increases sharply immediately before the reconnection. This is expected from the vortex line profiles for $t = t_{\mathrm{c}}$ in Figs.~\ref{fig:crowevolutionnondipolar} and \ref{fig:crowevolutiondipolar} since, at the respective vortex separation global minima where the reconnection is about to occur, the vortices' curvature is visibly large. In addition, large oscillations are observed at early times due to the initial transient in $\psi$. However, the behavior of $\langle\overline{\kappa(t)}\rangle$ in the intermediate stages is not universal and is instead dipole-dependent. Unless the dipole polarization is parallel to the $z$-axis, $\langle\overline{\kappa(t)}\rangle$ increases slowly until immediately before the reconnection. This rate of increase is higher for nonzero $\varepsilon_{\mathrm{dd}}$ than in the nondipolar limit since, as Figs~\ref{fig:crowevolutiondipolar} (a) and (b) demonstrate, the vortices are strongly perturbed along the axis of dipole polarization. This can be understood in terms of the $1 < q < 6$ modes contributing quite substantially to the Kelvin mode population in Figs.~\ref{fig:modeaveragedheatmaps} (b) and (c) during this time interval. Conversely, the mean vortex curvature in the $z$-polarized superfluid decreases as the dominant contributions in Fig.~\ref{fig:modeaveragedheatmaps} (d) are from the lowest mode, $q = 1$. 
	
\begin{figure}
		\centering
		\includegraphics[width=0.5\linewidth]{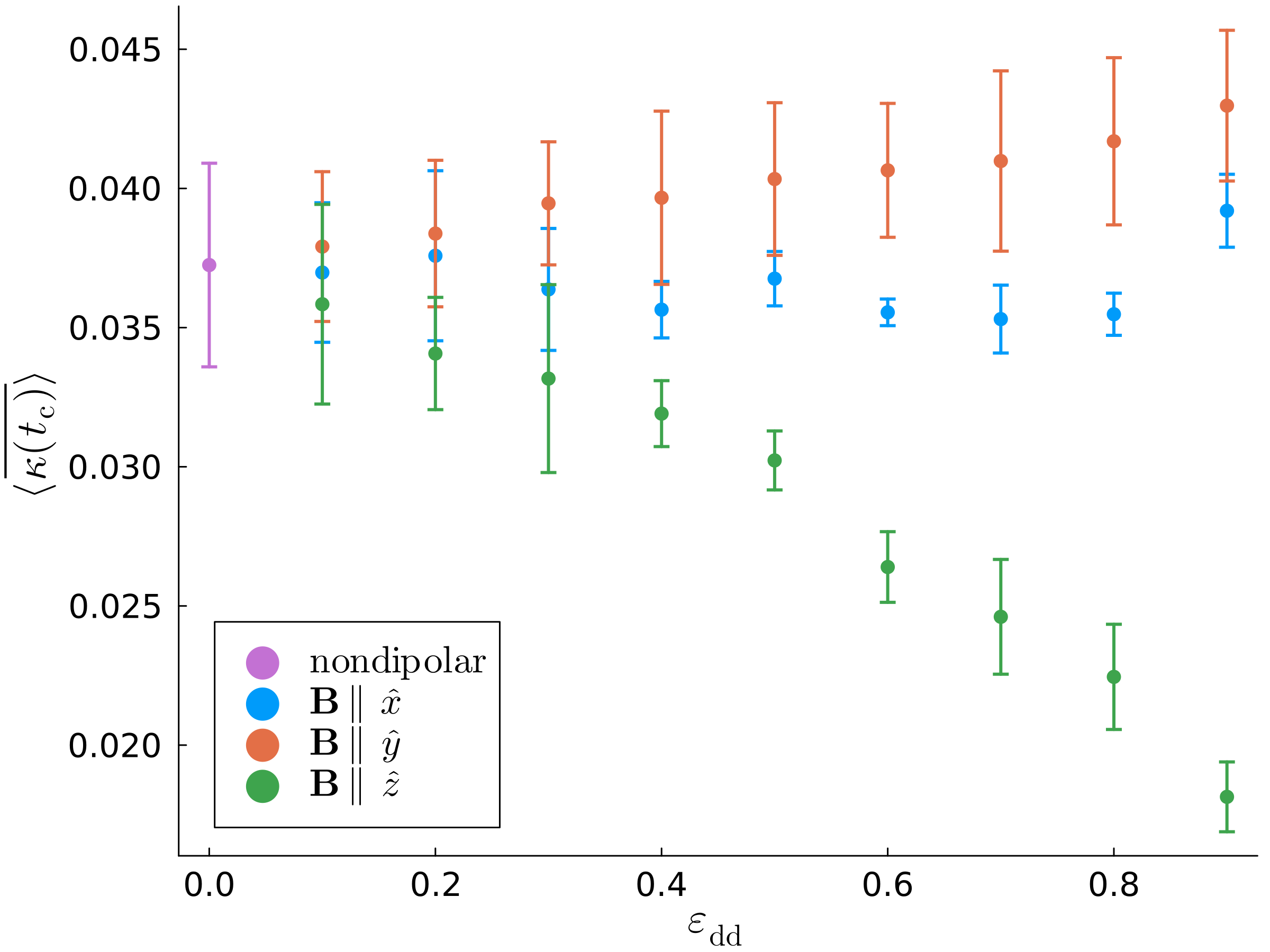}
		\caption{$\langle\overline{\kappa(t_{\mathrm{c}})}\rangle$, the mean curvature over the length of the two vortices ensemble-averaged over the set of vortex initial conditions at time $t = t_{\mathrm{c}}$. Each data series converges to the nondipolar value in the limit $\varepsilon_{\mathrm{dd}} = 0$, which we have represented via a distinct marker.}
		\label{fig:curvensavgtc}
	\end{figure}

	Further insight is afforded by comparing the global mean curvature for different dipolar regimes when $t = t_{\mathrm{c}}$. Figure \ref{fig:curvensavgtc} plots $\langle\overline{\kappa(t_{\mathrm{c}})}\rangle$ as a function of $\varepsilon_{\mathrm{dd}}$ with $\mathbf{B} \parallel \lbrace \hat{x}, \hat{y}, \hat{z} \rbrace$. While $\langle\overline{\kappa(t)}\rangle$ exhibits temporal fluctuations due to the randomness of the initial conditions, one finds that $\langle\kappa(t_{\mathrm{c}})\rangle$ is largely independent of $\varepsilon_{\mathrm{dd}}$ when the dipole moments are polarized along the $x$-axis, the translation axis of the vortices. As for polarizations along the $y$-axis, Fig.~\ref{fig:curvensavgtc} suggests that $\langle\kappa(t_{\mathrm{c}})\rangle$ generally increases with $\varepsilon_{\mathrm{dd}}$ but only slightly relative to the nondipolar value. This constrasts sharply to the case where the dipole moments are polarized along the $z$-axis, i.e. parallel to the vortex lines. Instead, $\langle\kappa(t_{\mathrm{c}})\rangle$ decreases monotonically for larger $\varepsilon_{\mathrm{dd}}$, such that the mean vortex curvature is not enhanced but suppressed by the DDI. The magnitude of this dependence of $\langle\kappa(t_{\mathrm{c}})\rangle$ on $\varepsilon_{\mathrm{dd}}$ is also larger than that when $\mathbf{B} \parallel \hat{y}$. In earlier studies of Kelvin waves on single, isolated vortices, this has been attributed to the effective interaction between the \textit{virtual dipole moments} that are induced in the vortex line~\cite{prl_100_24_240403_2008, njp_11_5_0155612_2009, pra_98_6_063620_2018}. Note that the virtual dipoles are antiparallel to the real dipole moments of the superfluid bulk. When $\mathbf{B} \parallel \hat{z}$ the contributions to the dipolar interaction energy arising from the interactions between the virtual dipole moments are minimized when the angle between any two virtual dipoles and their polarization axis is minimized, \textit{cf.} Eq.~\eqref{eq:ddireal}. This results in the vortex curvature being suppressed and also implies a lower wavenumber of the preferentially excited Kelvin mode when $\mathbf{B} \parallel \hat{z}$ than for the other polarizations in Fig.~\ref{fig:maxmodeeachB}. These effects are naturally more pronounced for larger values of $\varepsilon_{\mathrm{dd}}$, leading to the monotonic decrease of both $\langle\kappa(t_{\mathrm{c}})\rangle$ and $k_{\mathrm{c}}$ with respect to $\varepsilon_{\mathrm{dd}}$ in Figs.~\ref{fig:curvensavgtc} and \ref{fig:maxmodeeachB} (c), respectively. Furthermore, the inducement of virtual dipole moments in the vortex cores can explain the salient features of the vortices when $\mathbf{B}$ is parallel to $\hat{x}$ or $\hat{y}$. When $\mathbf{B} \parallel \hat{y}$, configurations of the virtual dipole moments along this axis are favored and, thus, perturbations are enhanced along this axis. While this results in the monotonicity of $\langle\kappa(t_{\mathrm{c}})\rangle$ in Fig.~\ref{fig:curvensavgtc}, it is also responsible for faster vortex reconnections than for the other dipole polarizations as observed in Fig.~\ref{fig:crowevolutiondipolar}. By extension, this is why the exponential mode growth factor $\sigma$ is largest when $\mathbf{B} \parallel \hat{y}$. Similarly, when $\mathbf{B} \parallel \hat{x}$, vortex line perturbations orthogonal to this axis are suppressed and thus the mean curvature arises mainly from perturbations parallel to the velocity of the vortex pair. While this results in the large transverse perturbations of the vortices in the $x$-$z$ plane in Fig.~\ref{fig:crowevolutiondipolar} (a), perturbations along $\hat{y}$ are greatly inhibited and thus the reconnection is greatly delayed. This dipole-mediated inhibition of the exponential growth of the dominant Kelvin mode when $\mathbf{B} \parallel \hat{x}$ is responsible for the smaller values of $\sigma$ in Fig.~\ref{fig:modeexponentensembledandmodegrowth} (b).
	
	\section*{Discussion and Conclusions}
	
	Through studying the evolution of a pair of antiparallel vortices in a dipolar superfluid when subjected to an initial transverse perturbation, we have found that the Crow instability exhibits a striking dependence on the magnetic dipole polarization and the relative strength of the DDI. When the axis of the applied magnetic field, $\mathbf{B}$, is parallel to the unperturbed vortex lines, the vortices' curvature is strongly suppressed by interaction between virtual dipole moments inside the vortex cores. This stiffening results in the dominant Kelvin modes excited by the Crow instability being of ever-longer wavelength as the relative dipolar interaction strength, $\varepsilon_{\mathrm{dd}}$, increases. These virtual dipole moments have the opposite effect when the dipole polarization is orthogonal to the initial vortex lines, with both the vortex curvature and the growth of higher Kelvin modes being enhanced by the DDI. While the Kelvin mode relative population for a dipole polarization parallel to the vortex separation is quite similar to its nondipolar counterpart, these modes induce larger vortex curvatures for higher $\varepsilon_{\mathrm{dd}}$ and, correspondingly, a more rapid onset of the Crow instability. However, for dipole polarizations parallel to the vortices' velocity, the Kelvin mode population is concentrated at a single mode, $q = 3$, in the large-$\varepsilon_{\mathrm{dd}}$ limit and the elongation of the vortex lines along the polarization axis inhibits the Crow instability when compared to the other two polarizations.
	
	The outcome that the Crow instability is strongly modulated by the dipole polarization suggests that other hydrodynamic instabilities might exhibit a directional dependence in a dipolar superfluid. Notably, the snake instability of dark solitons is suppressed when the dipole polarization is parallel to the solitonic phase slip~\cite{prl_101_21_210402_2008}. An intrinsic correspondence exists between the snake instability of dark solitons and the Crow instability of antiparallel vortices~\cite{pre_51_5_4479-4484_1995}, such that the analogue of this solitonic phase slip in our system is parallel to the $x$-axis. Given that our results demonstrate a suppression of the Crow instability for dipole polarizations along this axis, the twin predictions of instability suppression are probably not coincidental. We note that this prediction of the suppression of the snake instability was obtained via linear stability analysis of the excitation spectrum of a three-dimensional dark soliton~\cite{prl_101_21_210402_2008} and that similar methods have been used to analyze the stability of two antiparallel vortices against transverse perturbations in a nondipolar BEC~\cite{jphysa_34_47_10057-10066_2001}. It is thus likely that linear stability analysis of the stationary states of an antiparallel vortex pair in a dipolar superfluid would shed further insight into the dipolar Crow instability. We also note that Ref.~\cite{prl_101_21_210402_2008} did not consider dipole polarizations orthogonal to the solitonic phase slip. Given the contrasts we have observed for different dipole polarizations, we believe that it is prudent to also study the snake instability when the dipole polarization lies in the soliton plane. Another hydrodynamic instability arising from transverse perturbations is the Donnelly-Glaberson instability of vortex ensembles in rotating superfluids, where spontaneous radiation and amplification of Kelvin modes can lead to reconnections as well as the melting of a lattice of straight vortex lines into a tangled state~\cite{pra_79_3_033619_2009}. Noting our observations of the dipole-mediated enhancement or suppression of Kelvin mode growth in a pair of antiparallel vortices, a study of transverse perturbations of a much larger ensemble of vortices embedded in a dipolar superfluid might thus be warranted.

	\section*{Data availability}
	
	The datasets generated during and/or analysed during the current study are available at \url{https://doi.org/10.25405/data.ncl.30048364}.
	
	\section*{Methods}
	
	\subsection*{Propagating the dipolar Gross-Pitaevskii equation and extracting the vortex lines}
	
	Our results are obtained from numerically solving the dimensionless dGPE via the second-order split-step pseudospectral method~\cite{kinetrelatmod_6_1_1-135_2013}. The spatial domain has dimensions $\lbrace L_x, \,L_y, \,L_z\rbrace = \lbrace 100\xi, \,100\xi, \,200\xi\rbrace$ along the $x$, $y$- and $z$-axes, respectively, and is discretized evenly on a spatial grid of $256\,\times\,256\,\times\,512$ points. The use of discrete Fourier transforms in the pseudospectral scheme to calculate spatial derivatives results in periodic boundary conditions being imposed upon the solutions of Eq.~\eqref{eq:dgpescaled}. The contribution of the DDI to Eq.~\eqref{eq:dgpescaled} is also evaluated pseudospectrally via the convolution theorem and the analytical form of $V_{\mathrm{dd}}(\mathbf{r})$ in reciprocal space, $\widetilde{V}_{\mathrm{dd}}(\mathbf{q})$~\cite{repprogphys_72_12_126401_2009}:
	\begin{equation}
		\widetilde{V}_{\mathrm{dd}}(\mathbf{q}) = \left(\mathbf{B}\cdot\hat{\mathbf{q}}\right)^2 - \frac{1}{3}. \label{eq:ddifourier}
	\end{equation}
	The initial conditions of the dGPE are obtained by applying a Wick rotation to the dGPE such that $t \mapsto it$ and propagating it in \textit{imaginary time} with the desired initial phase of $\psi$ being imprinted continuously until $|\psi|$ converges. The numerical simulations of the evolution of the system are conducted by inverting the Wick rotation and evolving the dGPE in real time; for all of the parameters and initial conditions studied in our work, the imaginary and real timesteppers use timesteps of $dt = 0.01$ and $dt = 0.001$, respectively.
	
	The superfluid velocity corresponding to $\psi$ is given by $\mathbf{v} = \nabla S$ where $S = \nabla\left[\mathrm{Im}(\log\psi)\right]$, the phase of $S$, effectively plays the role of a superfluid velocity potential. Quantum vortices are characterized by both a quantized circulation~\cite{rmp_81_2_647-691_2009}, \textit{viz.}
	\begin{equation}
		\Gamma = \oint\,\mathrm{d}\mathbf{s}\cdot\mathbf{v} = 2\pi s\,:\,s\in\mathbb{Z}, \label{eq:circulation}
	\end{equation}
	around the vortex and a vanishing superfluid density, i.e. $|\psi|^2 = 0$, at the vortex core. To efficiently identify regions of the spatial domain where both properties are manifested, we compute the \textit{pseudovorticity}~\cite{jphysa_49_41_415502_2016},
	\begin{equation}
		\bm{\omega}_{\mathrm{ps}} = \frac{1}{2}\nabla\times\left(n\mathbf{v}\right) \equiv \nabla\mathrm{Re}[\psi] \times \nabla\mathrm{Im}[\psi]. \label{eq:pseudovorticity}
	\end{equation}
	This quantity is only nonzero in the vicinity of a quantum vortex core and has thus been used extensively in recent studies of superfluid vortex dynamics to locate and track vortices~\cite{prf_2_4_044701_2017, prx_7_2_021031_2017, pra_105_1_013304_2022, prr_5_4_043081_2023}. Vortex detection is therefore implemented as a two-stage process. First, the entire domain is divided into sets of discretized $x$-$y$, $y$-$z$ and $z$-$x$ planes; for example, each of the $x$-$y$ planes is defined such that $z$ equals one of its spatial gridpoints. On each plane, the circulation of $\mathbf{v}$ is computed via the trapezoidal rule and, given we expect only vortices of quanta $s = \pm 1$, a vortex is taken to lie near a gridpoint of the plane if $|\Gamma| > 6$. The subgrid domain between the identified gridpoint and the neighboring points is then divided into 10 intervals in each direction before the node of the vortex line is interpolated between each interval. This is achieved by employing the Newton-Raphson method to find the points where $|\psi|^2 < 10^{-5}$, with the search direction being weighted by the pseudovorticity to aid in its convergence. This procedure is applied every $500$ timesteps such that the interval between subsequent snapshots of the vortex profiles is $\Delta t = 0.5$.
	
	\subsection*{Superfluid phase: initial conditions}
	
	The correspondence between the phase of $\psi$ and the superfluid velocity, $\mathbf{v} = \nabla S$, allows us to use the classical result for the velocity potential of a pair of straight, antiparallel vortices in an incompressible fluid as the basis of the initial phase in the simulations. For vortices in the $x$-$y$ plane in a periodic domain with dimensions $x\in[0, L_x),\,y\in[0, L_y)$, this is given by~\cite{physfluidsa_3_5_835-844_1991, prl_112_14_145301_2014, pra_101_5_053601_2020}
	\begin{equation}
		S(x, y) = \sum_{p = -\infty}^{\infty}\,\sum_{n = 1}^2(-1)^n\biggl\lbrace\arctan\left[\tanh\left(\frac{\pi Y_n}{L_y} + p\pi\right)\cot\left(\frac{\pi X_n}{L_y}\right)\right] - \pi\Theta(X_n)\biggr\rbrace + \frac{2\pi(x_1 - x_2)y}{L_xL_y}, \label{eq:weissmcwilliamsphase}
	\end{equation}
	Here, the vortex with circulation $2\pi(-1)^n$ is located at $\mathbf{r}_n$ and $\mathbf{R}_n = \mathbf{r} - \mathbf{r}_n$. In practice, replacing the infinite sum over $p$ with a partial sum from $-P$ to $P$ results in a rapid convergence of $S$ for small positive $P$; our simulations fix $P = 11$ in line with our previous work~\cite{pra_109_6_063323_2024}.
	
	Let us define the following random perturbation,
	\begin{equation}
		\delta w(z) = \sum_{q = -Q}^{Q}\exp\left[i\pi(2\eta_q - 1) + \frac{2\pi iqz}{L_z}\right] - \exp\left[i\pi(2\eta_0 - 1)\right], \label{eq:kelvininitconds}
	\end{equation}
	representing a uniform population of the $40$ lowest-lying excitations along the $z$-axis, each with random phase $\eta_q$ drawn from the uniform distribution in the interval $[0, 1)$. As described in the Results, we populate only the lowest $40$ modes and, as such, $Q = 20$ throughput the article. The initial phase of $\psi$ in our simulations of the Crow instability is then given by Eq.~\eqref{eq:weissmcwilliamsphase} with $(x_n, y_n) \mapsto [x_n + (-1)^n\mathrm{Re}(\delta w), y_n + (-1)^n\mathrm{Im}(\delta w)]$.
	
	\subsection*{Kelvin mode amplitudes and vortex curvature}
	
	To determine the Kelvin mode amplitudes, $W_n(k_z, t)$, we represent the coordinates of the $n$th vortex line as $w_n(z) = x_n(z) + iy_n(z)$. Shifting the average of these coordinates over the $z$-axis to the origin of the $x$-$y$ plane yields the quantity $\tilde{w}_n(z, t) = w_n(z, t) - \langle w_n\rangle(t)$, where $\tilde{w}_n(z, t = 0) = (-1)^n\delta w(z)$ is the initial Kelvin wave perturbation given by Eq.~\eqref{eq:kelvininitconds}. The corresponding \textit{mode amplitudes}, $W_n(k_z, t)$, are the discrete Fourier transforms of $\tilde{w}_n(z, t)$ and the periodicity of the domain ensures a restriction of the mode wavenumber, $k_z$, such that $k_z = 2\pi q/L_z,\,q\in\mathbb{Z}$. We also note that, by construction, $W_n(k_z = 0, t) = 0$.
	
	The curvature of each vortex line can be computed directly from the mode amplitudes as follows. Noting that $\bm{\gamma}^{(n)}(s, t)$ is simply equivalent to $[x_n(s, t), y_n(s, t), s]$, the first and second derivatives in Eq.~\eqref{eq:curvaturedefn} can be found by multiplying $W(k_z, t)$ by $ik_z$ and $-k_z^2$, respectively, and transforming the product back to real space. Decomposing the result into its real and imaginary components yields the spatial derivatives with respect to $x$ and $y$, respectively. In practice, though, before inverting the discrete Fourier transform, it is necessary to apply a low-pass filter to minimize the effects of noise from the vortex detection procedure, as well as aliasing, that adversely affect the computation of higher-order derivatives. Through trial and error we have found that a simple low-pass filter that excises modes for which $|k_z| > 100\times 2\pi/L_z$ is sufficient.
	
	\bibliography{main.bbl}
	
	\section*{Acknowledgements}
	
	The authors acknowledge support from the Leverhulme Trust (Grant No. RPG-2021-108). N.G.P. also acknowledges support from the UK Engineering and Physical Sciences Research Council (Grant No.
	EP/T01573X/1). S.B.P. thanks Luca Galantucci for providing the Fortran package used herein to extract the vortex line profiles. This research made use of the Rocket High Performance Computing service at Newcastle University, and we thank Newcastle University for its provision.
	
	\section*{Author contributions statement}
	
	A.W.B. conceived the investigation, S.B.P. conducted the simulation and obtained the results, and all authors analysed the results. The manuscript was principally written by S.B.P. with contributions from N.G.P. and A.W.B., and all authors reviewed the manuscript.
	
	\section*{Declarations}
	
	\section*{Competing interests}
	
	The authors declare no competing interests.
	
\end{document}